\documentclass[usegraphicx]{mn2e}
\usepackage{amsmath}
\usepackage{ifthen}

\def \version {clean}

\ifthenelse{\equal{\version}{annotated}}
{
	\usepackage{ulem}			
	\newcommand{\deltext}[1]{\sout{#1}}	
	\newcommand{\newtext}[1]{{\bf #1}}	
	\newcommand{\comment}[1]{{\bf #1}}	
	\renewcommand{\em}{\it}
}{
	\newcommand{\deltext}[1]{}		
	\newcommand{\newtext}[1]{{#1}}		
	\newcommand{\comment}[1]{}		
}
	
\def\br {{\bf r}}
\def\bv {{\bf v}}
\def\sc {\rm}

\begin{document}

\title[Gravitational Redshifts in Clusters]{Measuring Gravitational Redshifts in Galaxy Clusters}
\author[Nick Kaiser]{Nick Kaiser \\
Institute for Astronomy, University of Hawaii}
\maketitle 

\begin{abstract} 
Wojtak {\it et al\/} have stacked 7,800 clusters from the SDSS survey in redshift space.  
They find a small net blue-shift for the cluster galaxies relative to
the brightest cluster galaxies, which agrees quite well
with the gravitational redshift predicted from GR.
Zhao {\it et al.\/} have pointed out that, in addition to the gravitational
redshift, one would expect to see transverse Doppler (TD) redshifts, so
$\langle \delta z \rangle = -\langle \Phi \rangle + \langle \beta^2 \rangle / 2$
with $\beta$ the 3D source velocity in units of $c$, and that these two effects
are generally of the same order.
Here we show that there are other corrections that are also of the same
order of magnitude.
The fact that we observe galaxies on our past light cone results in a bias such
that more of the galaxies observed are moving away from us in the frame of
the cluster than are moving towards us.  This causes the observed average
redshift to be 
$\langle \delta z \rangle = -\langle \Phi \rangle + \langle \beta^2 \rangle / 2 + \langle \beta_x^2 \rangle$,
with $\beta_x$ is the line of sight velocity.
That is if we average over galaxies with equal weight.
If the galaxies in each cluster are weighted by their fluence, or equivalently
if we do not resolve the moving sources, and make an average of the mean redshift
giving equal weight per photon, the observed redshift is
$\langle \delta z \rangle = -\langle \Phi \rangle - \langle \beta^2 \rangle / 2$,
so the kinematical effect is then opposite to the usual transverse Doppler effect.
In the WHH experiment, the weighting is a step-function because of the flux-limit for
inclusion in the spectroscopic sample and the result is different again,
and depends on the details of the luminosity function and the SEDs of the galaxies.
Including these effects substantially modifies the blue-shift profile.  
\deltext{We identify some potential biases in the dynamical analysis of stacked clusters.}
We show that in-fall and out-flow have very small effect over the relevant range of impact parameters
\newtext{but out-flow becomes significant and needs to be taken into account for
measurements on larger scales.}
\end{abstract}


\section{Introduction}

Wojtak, Hansen and Hjorth  (2011, hereafter WHH) have measured the gravitational redshift effect in
clusters of galaxies.  They stacked 7,800 massive clusters selected from the
GMBCG cluster sample (Hao, J., {\it et al.\/}, 2010) derived from from the SDSS DR7 
survey data (Abazajian {\it et al.\/}, 2009) in redshift space, using coordinates of the 
brightest cluster galaxy (BCG) as the origin.
They fit the cluster-frame redshift distributions, determined at a range of
impact parameters, to a linear
ramp to describe the foreground and background galaxies plus a quasi-Gaussian distribution to
describe the cluster, and find that the centres of the cluster components
have a small net blue-shift $\delta z \sim - 10\;{\rm km\;s}^{-1} / c$;
a remarkable achievement since the galaxy clusters have velocity dispersions
of order 600 km/s.  

A blue-shift would be expected in GR \newtext{(see e.g.\ Cappi, 1995)} since the light from the BCGs,
which are thought to reside close to the centres of clusters, will 
have climbed out from deeper in the cluster potential well than the light from
the majority of the galaxies, and the amplitude
of the effect appears to be broadly consistent with their estimates
of the gravitational redshift obtained using a mean cluster mass distribution determined from
the observed velocity dispersion.  

WHH suggested that the result is in conflict
with the predictions of TeVeS modified gravity theory (Bekenstein, 2004).  
However, agreement with the potential inferred from the kinematics of
non-relativistic particles like galaxies is expected in any metric
theory of gravity since both gravitational redshifts and particle motions
are determined by the time component of the metric;
what this type of measurement tests is the validity of the equivalence principle
(Will, 2006; Bekenstein and Sanders, 2012; Zhao {\it et al.\/}, 2012, hereafter ZPL).
This type of observation can, however, provide constraints on theories in which there
are long-range non-gravitational interactions between dark matter that augments gravity on cluster scales
(e.g. Gradwohl \& Frieman, 1992; Gubser \& Peebles, 2004; Farrar \& Rosen, 2007). 

WHH compare the \newtext{fractional} frequency shift with the estimate for $\langle \Phi \rangle_{r_\perp} / c^2$
where the averaging is along a line of sight with impact parameter $r_\perp$.
This would be appropriate if the light were emitted by non-inertial observers on a 
rigid, non-rotating, lattice in a state of rest with respect to the cluster.
It is also valid, to a good approximation, for observations of the redshift of
X-ray lines from heavy ions in the intra-cluster medium (Broadhurst \& Scannapieco, 2000).
But, as emphasised by ZPL, this is not correct when the light emanates
from galaxies that are in free-fall.  One way to obtain the
observed redshift in this situation is to use local Lorentz boosts to give the Doppler
shift between each emitting galaxy and its neighbouring lattice-based observer
living in the rest-frame of the cluster (though there are other constructions one could use --- see Bunn \& Hogg (2009) for an in-depth discussion).
If we set up coordinates such that the distant observer lives
at positive $x$, the energy of a photon in the emitting galaxy's frame relative
to the cluster rest-frame is $E_{\sc G} = \gamma (1 - \beta_x) E_{\sc RF}$ where
$\gamma = (1 - \beta^2)^{-1/2}$ and $\bbeta = {\bf v} / c$ with $\beta_x$ the
component towards the observer, so, up to second order in $\beta$
the redshift is $1 + z = E_{\sc G} / E_{\sc RF} = 1 - \beta_x + \beta^2 / 2$.
Adding the gravitational redshift difference yields the average redshift, given a phase space density
(PSD) for the galaxies $\rho(\br, \bbeta, t)$, of
\begin{equation}
\langle \delta z \rangle = 
\frac{\int d^3 r \int d^3 \beta \rho(\br, \bbeta, t) (- \beta_x + \beta^2 / 2 - \Phi/c^2)}{\int d^3 r \int d^3 \beta \rho(\br, \bbeta, t)}.
\end{equation}
Note that the redshifts here are all relative redshifts between observers and emitters in the
vicinity of the cluster, not the redshift actually observed; i.e.\ $1+z = (1 + Z_{\rm obs}) / (1 + Z_{\sc CL})$.

If the cluster is virialised, the PSD will be an even function of velocity so the mean 
of the line-of-sight velocity $\beta_x$ will
vanish, and one would conclude that the mean redshift difference is
\begin{equation}
\langle \delta z \rangle = \langle z_{\sc G} - z_{\sc BCG} \rangle = \langle \beta_{\sc G}^2 - \beta_{\sc BCG}^2 \rangle / 2
- \langle \Phi_{\sc G} - \Phi_{\sc BCG} \rangle / c^2
\end{equation}
where now, following ZPL, allowance is made for the fact that the BCG will, in general, not be at rest at the centre
of the cluster.  
Thus there is a positive contribution to the redshift, the transverse Doppler (TD) effect, that is opposite in sign to the
gravitational redshift (GR) effect for rest-frame emitters (it being assumed here that
the BCGs are on considerably lower energy orbits than  the general cluster population)
and, as emphasised by ZPL this effect will, quite
generally, be of the same order as the gravitational redshift for a bound system by virtue
of the virial theorem.

The point of this paper is to show that there are other corrections of
the same order of magnitude.
One arises from the fact that we observe the galaxies
on our past light cone and this causes a bias such that
we see more galaxies moving away from us than moving
towards us.  We show in \S2 that this gives an additional redshift $\langle \beta_x^2 \rangle$.

But that is only true if each source galaxy is weighted equally in the averaging.
If we apply any weighting based on galaxy luminosity then we also need to allow for
the special relativistic beaming effect.  In \S3 we show that if we do not resolve the
internal motions, but make an average that gives equal weight per observed photon,
the resulting redshift is just the opposite of the transverse Doppler effect.
That beaming and time-dilation would have an effect on gravitational redshift
measurements using X-ray observations was noted by Broadhurst \& Scannapieco (2000), but
in that application it is a much smaller effect so was ignored.

In the WHH experiment the weighting was a step-function imposed by the 
flux-limit for inclusion in the spectroscopic sample.  We calculate
the effect of this in \S4.  This turns out to be the dominant kinematic effect.

In \S5 we \deltext{first attempt to clarify come issues concerning dynamical
analysis of a composite cluster formed by stacking a heterogeneous
collection of clusters.  We then}
apply these results together with the observed velocity dispersion profile
to generate \newtext{revised} predictions for the net effect and compare with the observations.  

In \S6 we consider
the effect of infall and outflow, which we find to have very little impact on the measurements,
and in an appendix we develop the formalism for deriving, from numerical or analytical models, the
predicted distribution of observed redshifts in order to facilitate a more direct comparison
with the current and future observations.   

\section{Phase space density on the Past Light-Cone}

One might imagine that allowing
for the light travel time would be simply a matter of replacing $\rho(\br, \bbeta, t)$
in equation (1) by
$\rho(\br, \bbeta, t = x / c)$, in which case there would be no effect in the 
virialised region since for a stable, relaxed, system the PSD is independent
of time.  We are choosing the origin of time here to be the time the light
we observe left the center of the cluster. 

But this is not correct.  While the PSD is invariant under Lorentz boosts and
also along the trajectories of the particles, it has a non-trivial transformation
from rest-frame to light-cone (LC) coordinates:
\begin{equation}
\rho_{\sc LC}(\br, \bbeta) = (1 - \beta_x) \rho_{\sc RF}(\br, \bbeta).
\end{equation}
This means that the PSD for a virialised system viewed on the light cone
is not an even function of velocity but has a small asymmetry which results
in a non-vanishing of the mean of the line of sight component of the velocity.

One way to see how this arises is to consider taking a photograph of a swarm
of particles where, in any region of space, there are as many particles moving
towards us as away from us.  
The particles that we will see in a small cubical cell in space are not 
the same as the particles that occupy the cell at the moment the light
passes through the centre of the cell.  
As the past light cone of the event of our opening the shutter sweeps 
towards us through the cell it will overtake more particles that are
moving away from us than are moving towards us.  
The result is that more particles in the photograph
will have positive radial velocities than negative ones, \newtext{as was found in a slightly different context by Dunkel {\rm et al.\/}, 2009}.
More quantitatively, if we have a pair
of particles with the same $x$-component of velocity $\beta_x$ with separation
in the rest frame of $dx$ then on our past light-cone they have a separation
$dx_{\sc LC} = dx_{\sc RF} / (1 - \beta_x)$ and so the density (ordinary space
density or phase space density) on the light cone gains a factor $1 - \beta_x$.
Note that this is a purely Newtonian plus light-travel-time effect, and has
nothing to do with Lorentz-Fitzgerald length contraction which causes the
density of particles to depend on the state of motion of the observer.  It is
the same effect that causes a runner on a trail to meet more hikers coming
\deltext{towards her} \newtext{the other way} than going in the same direction.

This result can easily be verified in the case of
a toy model of a particle oscillating back and forth in a 1D parabolic
potential well $\Phi(x) = \omega^2 x^2 / 2$.  The trajectory is
$x(t) = a \cos(\omega t + \phi)$, where $\phi$ is the phase.  The
velocity in the rest-frame of the potential trough is 
$\beta = - (a \omega / c) \sin(\omega t + \phi)$ which, on
the light cone $t = x/c$ is $\beta = - (a \omega / c) \sin(\omega x/c + \phi) 
\simeq - (a \omega / c) \sin \phi - (a \omega / c)^2 \cos^2 \phi$.
The average of first term (over phase, or equivalently over time)
vanishes, but the second term is always negative and is just $-\beta^2$,
and the average agrees with $\langle \beta \rangle = \int dx \int d\beta
\rho(x,\beta) (1 - \beta) \beta / \int dx \int d\beta \rho(x,\beta) (1 - \beta) = -\langle \beta^2 \rangle$
with the rest frame PSD an even function of velocity.

For this toy model, and for particles uniformly distributed in phase,
 the PSD is zero except on a circle in
$(x',\beta)$ space (where $x' = x \omega / c$ is the dimensionless
displacement), and $\rho(x', \beta, t)$ vanishes except on a cylinder around the
$t$-axis.  When we slice this cylinder on the light cone, the particles also
live on a circle, but their density is non-uniform.

A parabolic 1-D potential is not very realistic, but the result is
quite general.  For particles orbiting in any static potential
well, the average of the instantaneous line of sight velocity, either
over time for one particle or over phase for a distribution, will
average to zero, but the observed velocity will contain an extra
term which is the light propagation time $x/c$ times the acceleration
of the particle, and the acceleration and position are anti-correlated in a gravitationally bound system,
so this does not average to zero.
Since the acceleration is the gradient of the potential, it is guaranteed
that the average of the line-of-sight velocity will be of the
same order as $\Phi / c$. 

WHH measured the mean redshift difference for galaxies at a range
of projected distances $r_\perp$ from the cluster center.  The appropriate
thing to compare with a PSD from a dynamical model or output of
a numerical simulation is, with suitable normalisation,
\begin{equation}
\begin{split}
\langle \delta z \rangle_{\br_\perp} & = \int dx \int d^3 \beta \rho_{\sc RF}(\br, \bbeta, t = x/c) 
(1 - \beta_x) \\
& \times (- \beta_x + \beta^2 / 2 - \Phi/c^2).
\end{split}
\end{equation}
In the virialised region, this gives
\begin{equation}
\begin{split}
\langle z - z_{\sc BCG} \rangle & = \langle \beta_x^2 - \beta_{x\sc BCG}^2 \rangle \\
& + \langle \beta^2 - \beta_{\sc BCG}^2 \rangle / 2 - \langle \Phi_{\sc G} - \Phi_{\sc BCG} \rangle / c^2.
\end{split}
\end{equation}
For isotropic orbits, the new term is 2/3 of the size of the TD effect and
is of the same sign.  Note that the asymmetry in the PSD applies to BCG
as well; BCG line of sight velocities will also be biased to be
positive with respect to the cluster centre of mass. 

\section{Unresolved sources}

The foregoing analysis assumes that the  redshift offsets are
determined from a catalogue of angular positions and redshifts,
thus effectively giving equal weight per galaxy.

But when we cannot resolve the sources, such as when we try to allow
for the kinematics of stars in BCGs, or, potentially, for low resolution
HI observations of clusters, we are averaging with equal weight per
observed {\em photon\/}, and this changes the effect.

Consider a source
that emits photons of fixed energy $E=E_0$ isotropically in its rest-frame
in a burst as it passes the origin of space moving at velocity $\beta$
along the $x$-axis.  Boosting the photon 4-momenta into the 
the `laboratory frame' (denoted below by primed coordinates) one finds
that a distant observer measures an energy $E(\mu') = E_0 / \gamma (1 - \beta \mu') = E_0 \gamma (1 + \beta \mu)$
where $\mu$ is the cosine of the angle between the $x$-axis and the photon direction.
Comparing the 3-momenta yields $\mu' = (\beta - \mu) / (1 + \beta \mu)$, and therefore
the Jacobian of the transformation from observed to source-frame solid
angles is $d\mu' / d \mu = 1 / \gamma^2 (1 + \beta \mu)^2 = (E_0 / E)^2$, and since
$n(\mu') d\mu' = n_0 d\mu$, the density of photons per unit solid angle is $n(\mu') = n_0 (E/E_0)^2 = n_0 / \gamma^2 (1 - \beta \mu')^2$.
This is the familiar relativistic beaming effect.
The energy is a function of lab-frame direction, and one finds that
the probability distribution for energy is $P(E) \propto n(\mu') d \mu' / dE(\mu')$
which is flat from $E = E_0 / \gamma (1 + \beta)$ to $E = E_0 / \gamma (1 - \beta)$,
and zero otherwise.

This is the probability distribution for random direction to the observer, or, equivalently,
the probability distribution for a single observer viewing radiation from 
sources at the origin moving in random directions.  It is also the same distribution
one would find for a particle oscillating back and forth in a box, or for the emission
from particles in a region of space if they are moving in randomly
oriented directions though all at the same speed.  The mean photon energy
is readily found to be $\langle E \rangle = \int dE\; E P(E) = \gamma E_0$; a result
that could have been anticipated since whatever rest-mass $\delta m_0$ the
source used to create the radiation has energy in the lab-frame $\gamma \delta m_0 c^2$.

For a distribution of velocities
we need to allow for the time-dilation effect: if the sources are identical
and all emitting photons at a fixed rate in the frame, the interval between
emission events in the observer-frame will be longer by a factor $\gamma$, so 
the number of photons observed per
unit time from sources with gamma factor $\gamma$ is $d\dot n(\lambda) \propto P(\gamma) d\gamma / \gamma$
and the average energy per photon is then
\begin{equation} 
\langle E \rangle = \frac{\int d\dot n \gamma E_0}{\int d\dot n} = 
\frac{\int d\gamma \gamma E_0 P(\gamma) / \gamma}{\int d\gamma P(\gamma) / \gamma}
= E_0 \langle \gamma^{-1} \rangle^{-1}.
\end{equation}
We see here that the received energy per unit time is just the sum of the power
of the sources, but the number of received photons per unit time in the observer frame has
a $1/\gamma$ dependence.

For unresolved sources then, the effect of the internal kinematics is to introduce
a blue-shift.  How does this square with the result obtained in the previous section where we found that the effect
for resolved sources was a red-shift that, once we allowed for light cone effects,
was actually larger than the transverse Doppler effect? To see that the 
two calculations are consistent, and obtain a useful check of the validity
of the light-cone effect, we now show that the resolved-source analysis
reproduces the result for unresolved sources, as of course it should, if
we introduce a weight per galaxy proportional to its fluence (number of 
photons per unit time per unit area at the detector).  It is sufficient
to consider identical isotropic emitters, for which the fluence is
$dn/dt \sim E n(E) \sim E^3$ where the extra factor of energy flux as compare to 
photon flux comes again from the transformation from intervals of time at the source and at the observer.

The number of photons per second received from such a source is proportional to
$1 / \gamma^3 (1 - \beta \mu')^3$ and therefore
the fluence weighted mean observed photon energy should be given by
\begin{equation}
\langle E \rangle = E_0 \frac{\int d\beta \; \beta^2 \rho_0(\beta) \int d \mu (1 - \beta \mu) \gamma^{-4} (1 - \beta \mu)^{-4}}
{\int d\beta \; \beta^2 \rho_0(\beta) \int d \mu (1 - \beta \mu) \gamma^{-3} (1 - \beta \mu)^{-3}}
\end{equation}
where the first factor of $1 - \beta \mu$ is the asymmetry of the observed phase space
distribution from the light-cone effect, 
and where we have assumed that the distribution of velocities is isotropic
and have dropped the prime.  The integrals are elementary, and we readily
find $\langle E \rangle = E_0 / \langle 1/\gamma \rangle$, fully consistent with the result obtained
above.  Without the light-cone $1 - \beta \mu$ term we would have obtained a different result.

For non-relativistic systems with $\beta^2 \ll 1$ the effect of the internal motions of emitters within
unresolved objects is to give a change of energy $\delta E/E_0 \simeq \langle \beta^2 \rangle /2$
which is a blue-shift and just opposite to the usual tranverse Doppler redshift.

\section{Surface Brightness Modulation}

In \S2 we implicitly assumed that all galaxies are observed and catalogued.  But in
reality the galaxies in the spectroscopic sample were selected according 
to their apparent luminosities.  As discussed in \S3, a galaxy's
apparent luminosity depends on its state of motion through the beaming effect
which changes the surface brightness and hence luminosity.   This results in a bias for the redshifts
of the cluster galaxies which are selected for redshift measurement as 
the surface brightness modulation means that
galaxies moving away from us in a given region of space have a higher limit on
their intrinsic luminosities than galaxies moving towards us. 
For the WHH experiment the effect is quite strong, and strongly
increasing with cluster redshift, since the flux limit ($r = 17.8$) is only about one magnitude fainter
than $M_*$ even at the minimum redshift limit, so the flux selection limit falls on the steep end
of the luminosity  function (LF) where small fractional changes in luminosity have a large effect on the
number of sources detected.

We then consider the effect on the BCGs.  For these the flux limit is irrelevant, but
there is a bias because velocities can change the ranking of the two brightest
galaxies.  This turns out to be almost independent of cluster redshift.

\subsection{Flux-Limited Galaxies}

\begin{figure}
\includegraphics[angle=-90,width=84mm]{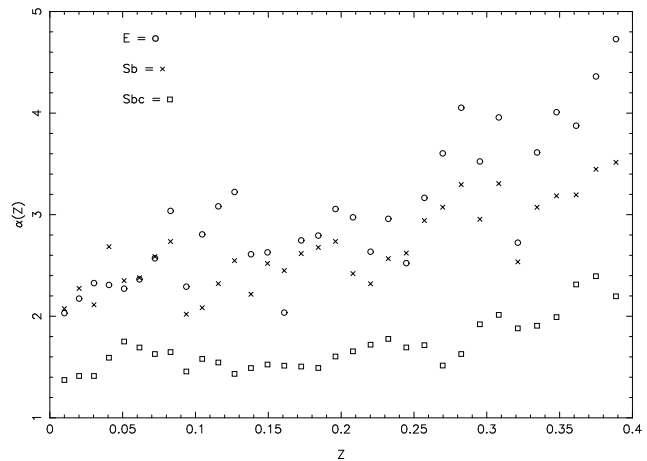}
\caption{Spectral index vs.\ redshift for representative galaxy types observed in Sloan r-band}
\label{alpha_z_fig}
\end{figure}

For low velocities $\beta \ll 1$ the fractional change of the fluence is $\Delta \dot  n / \dot n = 3 \beta_x$,
but to obtain the observed photon count,
we also need to allow for the change in frequency and the limits imposed by the broad-band
filter.  Combining these, the fractional change in apparent luminosity for a galaxy in
a cluster at redshift $Z$ is
$\Delta l / l = (3 + \alpha (Z)) \beta_x$,
where the effective spectral index, for a photon counting
detector with response curve $R(\lambda)$, is
$\alpha(Z) = - d \ln (\int d\lambda \lambda R(\lambda / (1 + Z)) f_\lambda) / d \ln (1 + Z)$.
\deltext{This depends on the details of the galaxy SED but
is found, using SEDs for E, S0 and Sb galaxies
from Coleman, Wu and Weedman (1980) to be 
close to 2 for galaxies at the relevant redshifts (see figure 1), reflecting the fact
that in the $r$-band, galaxies tend to have flat $f_\lambda$ curves.}
\newtext{This depends on the details of the galaxy spectral energy distributions (SEDs).
Baldry {\it et al.\/} (2002) provide an average SED for galaxies in the 2dfGRS which is
very similar to that of an Sb galaxy.  We expect the average for galaxies in
the main spectroscopic sample from SDSS to be very similar.  However, this is a luminosity
weighted average, whereas what is really needed is a number weighted average.
Since lower luminosity galaxies tend to be more active, this would be expected
to be bluer and therefore have a smaller spectral index.  A further complication
is that the WHH sample includes the luminous red galaxy (LRG) sample (Eisenstein {\it et al.\/}, 2001) that are redder - though these are relatively
few in number.  The effective spectral indices for representative galaxy types (E, Sb and Sbc)
from Coleman, Wu and Weedman (1980) and Kinney {\it et al.\/} (1996) are shown
in figure \ref{alpha_z_fig}. At the typical redshift for the sample the spectral index approximately
2.5 for the E and Sb SEDs and about 1.5 for the more active Sbc spectrum.  In what follows we
shall assume $\alpha \simeq 2.0$.}

The modulation of the number density of detectable
objects is given by the product of $\Delta l / l$ and the logarithmic derivative 
$\delta(Z) \equiv - d \ln n(>L_{\rm lim}(Z))/d \ln L$
which, as mentioned, is a strongly increasing function of redshift.  Ideally we would calculate this using
a luminosity function appropriate for the actual cluster galaxies used in the study, and this
will, in general, depend on the projected distance from the cluster center.
However, Hansen {\it el at.\/} (2009) have shown that while the
mix of red {\it vs.\/} blue galaxies changes strongly with radius, the
overall luminosity function does not vary much, and the parameters are
not very different from the field galaxy luminosity function, so we
will use the latter, as determined for SDSS
by Montero-Dorta \& Prada (2009), as a proxy.  Their estimate of the LF
obtained from the r-band magnitudes K-corrected to $Z=0.1$
has $M_* - 5 \log_{10} h = -20.7$ and faint end slope
of $\alpha = - 1.26$.  The resulting $d \ln n(>L)/d \ln L$,
computed using the flux limit $r = 17.77$ appropriate for
the SDSS spectroscopic sample used by WHH, is shown as
the dot-dash curve in figure 2.  \newtext{Note that we are ignoring
here the contribution from the LRG sample, which are selected 
according to a more complicated combination of colour
and luminosity cuts.}

Finally, we need the average of $-(3+\alpha) d \ln n(>L) / d \ln L$ over
the redshift distribution for the galaxies actually used in the experiment.
The 7,800 clusters used by WHH were selected
by applying a richness limit to the parent GMBCG catalogue (Hao, J., {\it et al.\/} 2010) that contains 
55,000 clusters extending to $Z = 0.55$.  These clusters were
derived from the SDSS photometric catalogue that is much deeper than the
spectroscopic catalogue.  Consequently, at the redshifts where the spectroscopically
selected galaxies live, this parent catalogue is essentially volume
limited for the clusters used, so the redshift distribution for the
cluster members used is essentially the same as that for the redshift
distribution for the entire spectroscopic sample, save for the
fact that the GMBCG catalogue has a lower redshift limit $Z_{\rm lim} = 0.1$, which
is very close to the redshift where $dN/dZ = Z^2 n(Z)$ peaks.  This is
the bell shaped curve in figure 2.  
Combining these we find $\langle d\ln n / d \ln L \rangle = 
\int dZ \; Z^2 n(Z) d\ln n / d \ln L / \int dZ \; Z^2 n(Z) \simeq -2.0$
with integration range $0.1 < Z < 0.4$, and the average 
$- \langle (3 + \alpha(Z)) d \ln n(>L) / d \ln L \rangle \simeq 10$.
This may be a slight overestimate, as the cluster catalogue is not precisely
volume limited and the actual $dN/dZ$ may lie a little below the solid
curve in figure 2 at the highest redshifts.

\begin{figure}
\includegraphics[angle=-90,width=84mm]{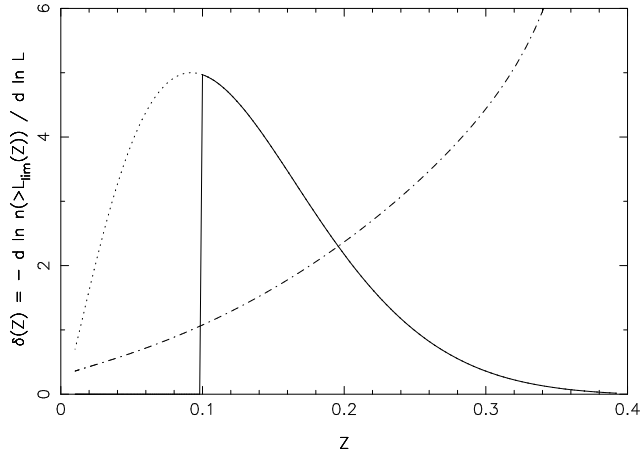}
\caption{The dot-dash curve is the logarithmic derivative of the comoving
density of objects above the luminosity limit as a function of redshift.
The bell-shaped curve is $dN/dZ = Z^2 n(Z)$ and the solid curve is that
truncated at the minimum redshift imposed by the parent cluster catalogue.
The mean  of the log-derivative, averaged over the redshift distribution
turns out to be $\simeq 2.0$.}
\label{dlnndlnL_fig}
\end{figure}

For the WHH experiment, the surface brightness modulation effect 
is considerably larger in amplitude than the transverse Doppler and light-cone effects, but has 
opposite sign.  For isotropic orbits the combination of the TD, LC and SB effects is 
\begin{equation}
\langle \delta z \rangle = (2.5 - \langle (3 + \alpha(Z)) \delta(Z) \rangle) 
\langle \beta_x^2 \rangle \simeq -7.5 \langle \beta_x^2 \rangle.
\end{equation}

\subsection{Effect on BCGs}

The TD and LC effects act on all galaxies, including the BCGs, in
the same way.  The SB effect is different; in the WHH analysis, only
clusters with at least 5 measured redshifts were used, so it is safe
to assume that the brightest cluster galaxies will be unaffected by the flux limit. 
However, for some small fraction of the clusters, the two brightest
galaxies will have magnitudes that are sufficiently close that
the effect of surface brightness modulation by the motions will be
enough to change their ranking, resulting in a bias.  In principle, this
effect could be eliminated by only using clusters where the difference
between the two brightest galaxies is sufficiently large that
the velocities cannot change the ranking.   

To analyse this, let the joint distribution
of difference of intrinsic magnitudes $m_{ab} = m_a - m_b$, and line-of-sight velocities $\beta_a$, $\beta_b$
for pairs of top two ranked cluster galaxies, in no particular order, be $P_0(m_{ab}, \beta_a, \beta_b)$.
This is a symmetric function of $m_{ab}$.

The velocities change the observed surface brightnesses of the galaxies, and hence the
difference of of observed magnitudes is $m'_{ab} = m_{ab} - \kappa (\beta_a - \beta_b)$,
where $\kappa \equiv (\ln(10) / 2.5) (3 + \alpha(Z))$, 
so the observed distribution is $P(m_{ab}, \beta_a, \beta_b) = P_0(m_{ab} - \kappa (\beta_a - \beta_b), \beta_a, \beta_b)$,
the Jacobian of the transformation from intrinsic to observed magnitude being unity.

The probability distribution for the velocity of the first ranked galaxy $\beta_1$ is then
\begin{equation}
\begin{split}
P(\beta_1)  & = \int_{-\infty}^0 d m_{ab} \int_{-\infty}^{\infty} d \beta_b P(m_{ab}, \beta_1, \beta_b) \\
& + \int_{0}^{\infty} d m_{ab} \int_{-\infty}^{\infty} d \beta_a P(m_{ab}, \beta_a, \beta_1) 
\end{split}
\end{equation}
i.e.\ the sum of the distribution function for $\beta_a$ if $m_{ab} < 0$ and the DF for $\beta_b$
if $m_{ab} > 0$.      

Making a Taylor expansion of $P(m_{ab}, \beta_a, \beta_b)$ with respect to $m_{ab}$ 
and performing the integrals yields
\begin{equation}
P(\beta_1) = P_0(\beta_1) - 2 \kappa \beta_1 P_0(m_{ab} = 0, \beta_1)
\end{equation}
which depends on the joint distribution of the intrinsic magnitude difference
and the velocity of one or other of the two brightest galaxies.
The mean redshift offset is then
\begin{equation}
\langle \delta z_{\rm BCG} \rangle 
= - \kappa \sigma_{ab}^2 P_0(m_{ab} = 0) / c^2
\end{equation}
where $\sigma_{ab}^2 \equiv \langle (\beta_a - \beta_b)^2 | m_{ab} = 0 \rangle$
is the variance of the relative velocity of the
two brightest galaxies given that they have similar magnitudes.  
This is something that is straightforward to measure from the data.  
It is reasonable to expect
that this is larger than (twice) the velocity variance for brightest
cluster galaxies.
Smith {\it et al.\/} (2010) have measured the distribution for
magnitude differences and find $P_0(m_{ab} = 0) \simeq 0.35$
so we then have 
\begin{equation}
\begin{split}
\langle \delta z_{\rm BCG} \rangle & \simeq 0.32 (3 + \alpha(Z)) \sigma_{ab}^2 / c^2 \\ 
& \simeq 1.9 \;{\rm km\;s}^{-1}/c (\sigma_{ab} / 600 {\rm km/s})^2.
\end{split}
\end{equation}
Note that the surface brightness boosting effect on BCGs does not have
the strong redshift dependence that is expected for the flux-selected
galaxies.

\section{Revised Redshift Profile Prediction}

We can now make a prediction for the combined GR+TD+LC+SB effect as a function
of radius using the observed velocity dispersion data provided by WHH.
\deltext{We start with a discussion of dynamical analysis of a composite, or stacked,
cluster and how this differs from the analysis of single cluster.}
We first review the relevant properties of the BCGs; both their cluster-centric 
kinematics and their halo properties.
We then attempt to combine all of the effects discussed above to
predict the expected profile of the redshift offset \newtext{and its constituent components.
We caution at the outset that there are considerable uncertainties
in many of the critical inputs that feed through into the prediction.}

\subsection{Motion of BCGs Within Clusters}

The analysis of WHH relies on the assumption that,
in an average sense, the BCGs used as the origin of coordinates in velocity and angle space
are a relatively cold population, velocity-wise, compared to the other galaxies and are therefore
orbiting close to the potential minimum.   There are good
theoretical grounds for believing that the BCGs will indeed be colder than
the general population, but understanding in detail just how cold they are
is important here for two reasons: first because \newtext{both the gravitational redshift and} the kinematically sourced
effects depend on the velocity dispersions of the
BCGs \newtext{which, following WHH,  must be subtracted in quadrature from the observed dispersion profile,}  and second because it can inform us
to what extent the mean density profile around the BCG is in fact likely
to depart from the idealised NFW model predictions \newtext{if the BCGs do not reside
precisely at the minimum of the potential}.

WHH assume $\sigma_{\rm BCG} = 0.35 \sigma_{\rm obs}$, 
citing Skibba {\it et al.\/} (2011), in which case the effect
on estimates of e.g.\ the TD and LC effects is quite small.
But this may be a bit low. Skibba {\it et al.\/} found that
the velocity dispersion for the {\em central\/} galaxies in the
clusters were $\sigma_{\rm cen} \simeq 0.5 \sigma_{\rm CL}$ which is
a lot larger, but not directly measuring the same thing since
they also found that about 30\% of the time, the central
galaxy was not in fact the brightest galaxy in the cluster.

Coziol {\it et al.\/} (2009) have measured the distribution of
BCG motions directly and
find that $\langle |v_{\rm BCG}| \rangle / \sigma_{\rm CL} \simeq 0.40 \pm 0.04$
for clusters of Abell richness class R=1.  These clusters have
mean dispersion $\sigma = 651$\;km/s, a little higher than for the composite cluster here.
The mean dispersion for R=0 is $\sigma = 539$\;km/s, for which they
find $\langle |v_{\rm BCG}| \rangle / \sigma_{\rm CL} \simeq 0.43 \pm 0.03$
so the appropriate value for the sample here is around 0.42.

For a Gaussian distribution, $\langle |v_{\rm BCG}| \rangle = \sqrt{2/\pi} \sigma_{\rm BCG}$
so this suggests $\sigma_{\rm BCG} = 0.53 \sigma_{\rm  CL}$ which
is again considerably larger than the value adopted by WHH and consistent
with what Skibba {\it et al.\/} found for the central cluster galaxies.

\newtext{There is clearly considerable uncertainty in the true value of $\sigma_{\rm BCG}$. 
In what follows we will assume $\sigma_{\rm BCG} = 0.50 \sigma_{\rm  CL}$.}
If we define $\alpha \equiv \sigma_{\rm BCG}^2 / \sigma_{\rm CL}^2$ then
we have $\sigma_{\rm CL}^2 = \sigma_{\rm obs}^2 / (1 + \alpha)$ and
$\sigma_{\rm BCG}^2 = (\alpha / (1 + \alpha)) \sigma_{\rm obs}^2$.
For $\alpha = 0.25$ \deltext{, as suggested by the observations,} and $\sigma_{\rm obs} = 620 \;{\rm km/s}$
we have $\sigma_{\rm BCG} = 277 \;{\rm km/s}$ and $\sigma_{\rm CL} = 555 \;{\rm km/s}$.
With this value the differential TD and LC
effects are reduced to about 60\% of what one would
expect in the limit that the BCGs lie at rest at the minimum of the cluster potential.
The SB effect, as we show below, is somewhat less affected.

We can also estimate the reduction in the gravitational redshift,
assuming that the clusters in which the BCGs live do indeed have NFW profiles. 
Vanishing of the second time derivative of the
moment of inertia $I = \sum r^2$ tells us that $\langle |\dot r|^2 \rangle = 3 \sigma_{\rm BCG}^2 = \langle r |\nabla \Phi| \rangle$.
In the inner parts of the NFW profile, the potential increases linearly
with radius, so consequently we have $\langle r |\nabla \Phi| \rangle = \langle \Phi \rangle$
so the predicted gravitational blue-shift for the hot population relative
to the colder BCG population is decreased by $\delta z = 3 \sigma_{\rm BCG}^2 / c^2 \simeq 0.9 (\sigma_{\rm BCG} / 300 \;{\rm km/s})^2 \;{\rm km\;s}^{-1}/c$.
But their motions will also give them TD and LC red-shifts that are $\delta z \simeq 2.5 \sigma_{\rm BCG}^2 / c^2$
which largely counteracts the change in the GR effect, and from \S4.2 the SB effect on the BCGs
gives them a blue shift which we have estimated to be about $c \delta z = -1.9$ km/s.

Finally, one should allow for the fact that the light we see from the galaxies will have suffered
gravitational redshift escaping the halos of the galaxies, and that the starlight will also be
affected by stellar motions as described above in \S3. This is most important for
the BCGs.  

\subsection{BCG Internal Kinematics and Halo Properties}

Regarding the GR effect,
the stellar velocity dispersion in BCGs is typically $\sigma_* \sim $250 km/s (e.g.\ Bernardi {\it et al.\/}, 2007); much
larger than that of the run-of-the-mill galaxies, and quite comparable to the
motion of the BCGs in the cluster halo.  The BCGs are unresolved, so we can use the
result of \S3 to predict the kinematically sourced blue-shift $\delta z \sim -(3/2) (\sigma_*/c)^2 \sim -0.3 \;{\rm km\;s}^{-1}/c$,
which is quite small.  If they have flat rotation
curve halos, for which $\Phi = \Phi_0 \ln r$ then, for isotropic orbits, $\Phi_0 = 2 \sigma^2_{DM}$
while vanishing of $\ddot I$ for the stars requires $\Phi_0 = 3 \sigma_*^2$.
The gravitational redshift is therefore $\delta z = 3 (\sigma_*^2 / c^2) \langle \ln(r_{\rm halo} / r_*) \rangle
= 0.63 \;{\rm km\;s}^{-1}/c (\sigma_* / 250 {\rm km/s})^2 \langle \ln(r_{\rm halo} / r_*) \rangle $.
The problem here is determining the logarithm since we need to know the
size of the BCG halo (as distinct from the cluster halo).  This could be
determined by galaxy-galaxy lensing, and also could in principle be determined
from simulations.  

A rough estimate can be obtained from tidal considerations:  The
BCG halo density is $\rho_{\rm halo} \sim 3 \sigma_*^2 / 4 \pi G r_{\rm halo}^2$ while
the density of the cluster is $\rho_{\rm CL} \sim 2 \sigma_{\rm CL}^2 / 4 \pi G r^2$
where now $r$ is the typical cluster-centric distance to the BCG.  If this is a
few hundred kpc then the tidal constraint that $\rho_{\rm halo} \ge \rho_{\rm CL}$
says that $r_{\rm halo}$ can't be bigger than about 1/3 of this.
The scale lengths for BCGs are typically 10kpc, so this would suggest that the
logarithm is approximately 2.5.  If the halos are really this large, the effect of the motion of the stars
$\delta z \simeq - 3 \sigma_*^2 / 2 c^2$ is a small correction, and we have
a net redshift $\delta z \simeq 1.25 \;{\rm km\;s}^{-1}/c $.

\subsection{Combined Prediction}

We will proceed in two steps.  We bootstrap off the estimate of the difference
in potential between the BCG and the innermost point using the WHH stacked NFW model
method.  The innermost data point lies at $r_\perp = 0.6$Mpc where the assumptions of virial
equilibrium are likely to be well obeyed.  We then extrapolate to larger impact 
parameters assuming galaxies trace the mass and using the cluster-galaxy cross-correlation
function to get the appropriate ensemble average mass profile.

The stacked NFW model appears to provide a good fit to the data within
1.2Mpc (WHH supplementary figure 2) and yields a potential for galaxies
at impact parameter $0.6$Mpc corresponding to $\delta z_{\rm GR} \simeq -5.0{\rm km /s}/c$.
The finite BCG velocities will reduce the
gravitational potential difference by about $0.9 {\rm km /s}/c$ but the
BGH halo potential increases it by an estimated $1.6 {\rm km /s}/c$.
The net result is a potential difference of $\delta z_{\rm GR}(0.6{\rm Mpc}) \simeq -5.7 {\rm km /s}/c$

We now need include the kinematic effects.  The observed velocity dispersion at
this impact parameter is $\sigma_{\rm obs} \simeq 620$\;km/s, so with $\sigma_{\rm BCG} = 0.5 \sigma_{\rm CL}$
the LC and TD effects are $\delta z_{\rm TD+LC} \simeq (3/5) 2.5 \sigma_{\rm obs}^2 / c^2 = + 1.9 \;{\rm km \;s}^{-1}/c$.
The SB effect for the non-BCGs is 
$\delta z_{\rm SB} \simeq -10.0 \sigma_{\rm CL}^2 / c^2 \simeq -10.3 \;{\rm km \;s}^{-1}/c$
and the SB effect on the BCGs we have estimated to be about $+1.9 \;{\rm km\;s}^{-1}/c$, and finally the 
kinematic blue-shift for the stars in the BCG gives $+0.3 \;{\rm km \;s}^{-1}/c$ for
a net kinematic effect $\delta z_{\rm TD+LC+SB}(0.6{\rm Mpc}) \simeq -6.1\;{\rm km \;s}^{-1}/c$
for a grand total $\delta z_{\rm GR+TD+LC+SB}(0.6{\rm Mpc}) \simeq -11.8\;{\rm km \;s}^{-1}/c$.  
whereas the observed value is $\delta z \simeq -2.6 \;{\rm  km\;s}^{-1}/c$.  The uncertainty
on this point is approximately $6 \;{\rm  km\;s}^{-1}/c$ so this would appear to be discrepant,
but only at about the 1.5-sigma level.

The NFW model predicts $\delta z \simeq -10 \;{\rm  km\;s}^{-1}/c$ for the outer
measurements $r \simeq 3.3, 5.3$Mpc, and the measurements straddle this value.
While this model may provide a reasonable description for isolated
clusters in the virialised domain, it is not at all clear that it is appropriate to
describe the composite cluster being studied here.  Tavio {\it et al.\/} (2008)
have claimed that beyond the virial radius the density in numerical LCDM simulations
actually falls off like $\rho \sim 1/r$
rather than the $\rho \sim 1/r^3$ asymptote for the NFW profile, and the
extended peculiar in-fall velocities found by Cecccarelli {\it et al.\/} (2011)
also argue for shallow cluster profiles, but it is
not clear that these results are widely accepted.  

An alternative, and possibly more reliable, approach is to
assume that galaxies trace the mass reasonably well, in which case the density profile of the stacked
cluster has the same shape as the cluster-galaxy cross correlation function 
(e.g.\ Lilje \& Efstathiou, 1988; Croft {\it et al\/}, 1997).
This has a power-law dependence $\rho \sim r^{-\gamma}$ with $\gamma \simeq 2.2$, i.e.\ intermediate
between the NFW and Tavio {\it et al.\/} model predictions.

For space density $\rho(r) = \rho_0 (r/r_0)^{-\gamma}$,
where $r_0$ is an arbitrary fiducial radius,
the potential is $\Phi(r) = \Phi_0 (r/r_0)^{2-\gamma}$
and the 1-D velocity disperson, for isotropic orbits, is $\sigma^2(r) = \sigma^2_0 (r/r_0)^{2-\gamma}$
with $\Phi_0 = 2 ((1 - \gamma)/(2 - \gamma)) \sigma_0^2$.   

\begin{figure}
\includegraphics[angle=-90,width=84mm]{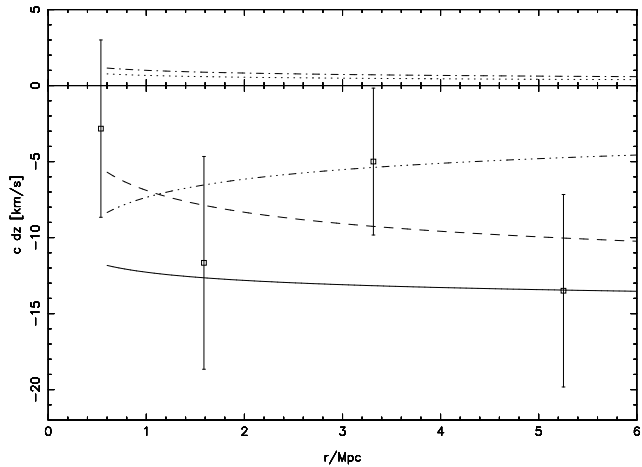}
\caption{Data points from figure 2 of WHH and prediction based on mass-traces-light
cluster halo profile and measured velocity dispersions as described in the main text.
The dashed line is the gravitational redshift prediction, which is similar
to the WHH model prediction.
The dot-dash line is the transverse Doppler effect.  The dotted line is the LC
effect.  The triple dot-dash line is the surface brightness effect.
The solid curve is the combined effect.}
\label{whh+pred_fig}
\end{figure}

The projected velocity dispersion measured is related to the 3-D velocity dispersion
by $\sigma^2(r_\perp) / \sigma^2(r) = \int dy\; y^{2-\gamma} (1 + y^2)^{-\gamma / 2}$
but the projected potential is related to the 3-D potential in the same way, so
the projected quantities are related by $\Phi(r_\perp) = 2 ((1-\gamma)/(2 - \gamma)) \sigma^2(r_\perp)$.
\comment{Fixed typo.}
This is the potential relative to infinity.  The difference in projected potential between
two projected radii $r_1$ and $r_2$ is
$\Phi(r_2) - \Phi(r_1) \simeq 12 \sigma^2(r_1) (1 - (r_1/r_2)^{0.2})$ 
\comment{Fixed typo.}
for $\gamma = 2.2$.
The resulting GR effect is shown as the dashed line in figure \ref{whh+pred_fig} and
is actually quite similar to the shape of the profile for the WHH NFW composite model.

The FWHM of the bell-shaped velocity distributions in WHH figure 1 appear to
decrease by about 15\% between the inner-bin and the outer points.  This is
reasonably consistent with the expected $\sigma^2 \propto r^{-0.2}$ trend predicted
if galaxies trace mass, but this is perhaps fortuitous since the outer
points are well outside the virial radius.  Regardless of whether the galaxies
at large radius are equilibrated or not, we can use the change in the observed velocity dispersion
with radius to obtain the differential TD+LC+SB effect which is shown,
added to the GR effect, as the solid line in figure \ref{whh+pred_fig}.
The kinematic effects flatten out the predicted profile, so the prediction
is quite different from the gravitational redshift alone.

The situation is clearly rather complicated, especially when using BGCs as the
origin of coordinates since the effects depend on things like the
relative velocities of the top ranked pair of cluster galaxies, and on the BCG halo properties,
that are quite poorly known.  However, those factors only influence the prediction
for the innermost data point.  The empirically based theoretical prediction for the
profile of the redshift offset for the hot population as a function of
impact parameter at $r_\perp > 0.6 {\rm Mpc}$ 
is the most robust; if galaxies are reasonable tracers of
the mass then profile should be very flat, quite unlike the GR effect from a NFW profile.
The predicted GR and total effects are shown in figure 3.
However, this analysis ignores the effect
of secular infall and out-flow which we consider next.

\section{Effect of Infall and Outflow}

The discussion so far has focused mostly on the stable, virialised regions.
Clusters, however, are evolving structures and the mass within any fixed physical
radius $M(r)$ will in general be changing.  Outside of the virial
radius (generally considered, inspired by the spherical
collapse model, to be the radius within which the mean
enclosed mass density is $3 \pi / G t^2$)
we expect to see net
infall, and the enclosed mass at those radii will be increasing with time, while at still larger
radii there will be outflow tending asymptotically toward the Hubble
flow.  In the spherical collapse model the transition from inflow to outflow takes place at
the turnaround radius where the mean enclosed mass density is ${\overline \rho}_{\rm t} = 3 \pi / 32 G t^2$.
This is for a matter dominated Universe; allowing for a cosmological constant makes only
a small change (Lokas \& Hoffman, 2001).

For the empirically motivated $\rho = \rho_0 (r/ r_0)^{-\gamma}$ model the 
mean enclosed mass is 
${\overline \rho}(r) = 3 (\gamma - 1) (2 \pi G)^{-1} \sigma_0^2 r_0^{\gamma - 2} r^{-\gamma}$
and the nominal virial radius is 
$r_{\rm vir} = ((\gamma - 1) \sigma_0^2 r_0^{\gamma - 2} t^2 / 2 \pi^2)^{1/\gamma} \simeq 1.8 {\rm Mpc}$
using $\gamma = 2.2$, $r_0 = 1$Mpc, $\sigma_0 = 545 \;{\rm km/s}$ and $t \simeq 1/H = 1/(70 \;{\rm km s}^{-1}/{\rm Mpc})$ and
turnaround is at $r_{\rm t} \simeq 8.7$Mpc.

In the centres of clusters there may be softening
of the cores which would reduce the enclosed mass and would have an associated outflow.

In any single cluster, the density may be changing rapidly --- on the local
dynamical timescale --- especially during mergers and as clumps rain in, 
but for a composite cluster such as considered here
these rapid changes will average out and the mass can only change on
a cosmological timescale: $\dot M \sim H M$.  
For power law profile with $\gamma \simeq 2$ 
$M \simeq 4 \pi \rho r3$ and $\dot M \simeq 4 \pi \rho r^2 \overline v$,
where $\overline v$ is the mean infall velocity, by continuity.  So if
$\dot M(r) = \alpha(r) H M(r)$, with, by the above argument, $\alpha(r)$ of order unity, 
then ${\overline \bv}(\br) = \alpha(r) H \br$.

This secular flow can generate a net offset for the redshifts in two ways.
First, and most importantly for clusters at $Z \ll 1$, along any line of sight we observe galaxies
that lie in a cone that will be wider at the back of the cluster.  At low
redshift this means there will be more galaxies observed at the back than
the front in an intrinsically symmetric cluster.  But we also need to allow for
the countervailing bias caused by the fact that the
more distant galaxies will be fainter which, as we have seen above, overwhelms the
effect of the change of volume in the relevant range of redshifts.
These geometric and flux limit effects, whose effects on the
foreground and background galaxies was discussed by Kim and Croft (2004),
modulate the density per unit line of sight distance linearly with
distance.  The real flux limited galaxies observed along 
cones behave like particles with no flux selection observed in cylinders with
a phase space density $\rho'(\br, \bbeta) = \rho(\br, \bbeta) (1 + 2 H x (\delta(Z) - 1)/cZ)$.

We can try to use this to estimate the redshift offset as 
$\int dx \int d^3 \beta \rho'(\br, \bbeta) \beta_x / \int dx \int d^3 \beta \rho(\br, \bbeta)$.
Performing the integrals over velocity this is
\begin{equation}
\langle \beta_x \rangle_{r_\perp} = \int dx \; \left(1 + \frac{2 H x}{cZ}(\delta(Z) - 1)\right) \rho(\br) {\overline \beta}_x(\br) 
/ \int dx \; \rho(\br)
\end{equation}
or, with $ {\overline \beta}_x(\br) = \alpha(r) H x / c$ and an assumed $\sim 1/r^2$ density profile,
\begin{equation}
\langle \beta_x \rangle_{r_\perp} = \frac{2 H^2}{c^2Z}(\delta(Z) - 1)) 
\int dx \; \frac{\alpha(r) x^2}{r^2} / \int \frac{dx}{r^2}
\end{equation}
where $r = \sqrt{r_\perp^2 + x^2}$.  This is rather messy and,
owing to the presence of the factor $\alpha(r)$, model dependent.  But
we can note the following: if we work at $r_\perp \sim 1 {\rm Mpc}/h$ say,
the integral in the denominator will be $\sim 1/r_\perp$ while the
contribution to the integral in the numerator from $-r_\perp < x < r_\perp$ will be $\sim r_\perp \alpha(r_\perp)$, 
so we get a partial contribution to $\langle \beta_x \rangle_{r_\perp} \sim H^2 r_\perp^2 / c^2 Z \simeq 5 \times 10^{-7} Z_{0.2}^{-1}$,
which is very small; corresponding to a physical velocity of only 0.16 km/s.  If we extend the range of integration
beyond $|x| < r_\perp$ this will increase, but not by a very large factor, since $\alpha(r)$ will start its decrease towards
zero at the turnaround radius.  Extending the range of integration still further the average decreases  since
$\alpha$ now has changed sign.  Ultimately, the value of this integral will become large, and even more
so beyond the $\sim 10 {\rm Mpc}/h$ scale of the cluster-galaxy cross-correlation function, but
this would not be seen as a shift of the bell-shaped enhancement of the redshift distribution.

The secular infall or outflow can also couple to the time rate of change of the
phase space distribution function, which will also be changing on a 
cosmological timescale (both the density and the width of the velocity
distribution will, in general, be varying).  This will result in
a formally similar contribution to the redshift offset, but without the
factor $1/Z$, so for low-redshift clusters this will be still smaller.

At larger impact parameter the effect of outflow is more interesting.
In figure 3 we show the line-of-sight velocity distribution for a simple,
but plausible, model for cluster density and velocity structure in the out-flow region.  
The key assumption here is that outside of turnaround the radial
velocity is $v \simeq H(r - r_{\rm t})$ and is supported
but the analysis of peculiar velocity profiles numerical simulations (Ceccarelli, {\it et al.\/}, 2011).
The amplitude of the effect has been exaggerated in figure 4 by a factor 10 for clarity.  For the average of $2 H (\delta(Z) - 1) /c Z$
over the distribution of galaxy redshifts shown in figure 2 is approximately 0.0023/Mpc.
Scaling the the mean velocity from the model appropriately
yields an expected blue-shift driven by the out-flow of about $\delta z \simeq -3.0 \;{\rm km\;s}^{-1}/c$
at impact parameter $r_\perp \simeq 10 {\rm Mpc}$.  Over the range of impact parameters explored by
WHH the effect of infall and outflow is very small.

\begin{figure}
\includegraphics[angle=-90,width=84mm]{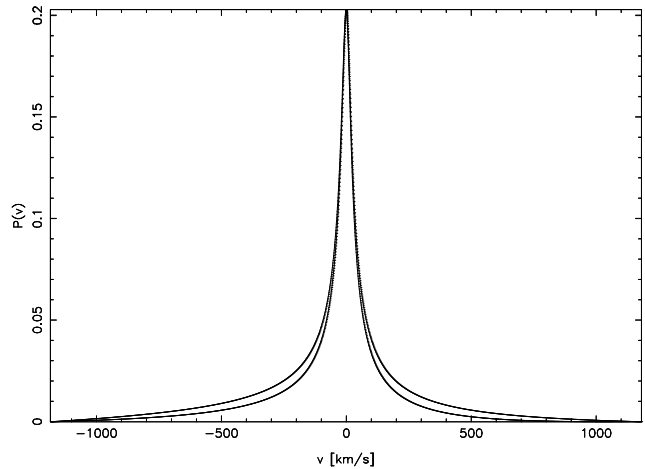}
\caption{Distribution of line of sight velocities at impact parameter $r_\perp = 10$Mpc
for a simple model where $\xi_{\rm cg} = (r/r_0)^{-2}$ with
correlation length $r_0 = 10$Mpc and turnaround radius $r_{\rm t} = 8.7$Mpc and where the velocity is
$v = H (r - r_{\rm t})$. 
A linear ramp determined from the outermost points has been subtracted.
These were generated using equation 13 with parameter $2 H (\delta(Z) - 1) / c Z = 0$ (thin line)
and 0.023/Mpc (thick line).  This is highly exaggerated. For clusters at $Z=0.2$,
or averaged over the distribution of galaxy redshifts, this parameter is
$\simeq 0.0023$.  Scaling the mean velocity appropriately gives $\delta z \simeq -3.0 \;{\rm km\;s}^{-1}/c$.
This is for cold spherical outflow.  In reality this will be convolved with
a broad quasi-Gaussian distribution of random velocities from local substructures
but the shift of the centroid will be essentially the mean of the distribution shown here.}
\label{outflow_fig}
\end{figure}

\section{Discussion}

We have shown that, in addition to the transverse Doppler effect, there are
additional factors that need to be taken into account
in interpreting the measurement of the offset of the net blue-shift of the cluster galaxies relative
to the central brightest cluster galaxy.  These are
straightforward to estimate from the measured line-of-sight velocities and
can therefore be subtracted from the measured blue-shift.  The TD effect
is a little more difficult to estimate, as it depends to some degree on
the velocity dispersion anisotropy, but as it is a relatively small
effect for the WHH experiment little error is made if we assume isotropic
orbits.  We have also shown that the redshift offset for unresolved sources
is different again; the kinematically sourced effect is a blue-shift
that is just the opposite of the standard transverse Doppler term.

We have applied these results to the WHH measurement. We have used an empirically
motivated model for the composite cluster halo mass density profile together with the observed velocity dispersions
to predict the net redshift offset. The largest correction comes from the surface brightness
modulation effect.  This is roughly equal to the GR effect at small
impact parameters, and, since the velocity dispersion is falling with radius, this flattens out
the blue-shift profile.  The result, it has to be admitted, does not seem
to agree as well with the data as the GR prediction alone.

The current data do not place particularly strong constraints on theories
that invoke long-range non-gravitational interactions in the dark sector.
However, the observational situation has already improved substantially with nearly three
times as many galaxy redshifts obtained by the Sloan telescope once one
includes the extensions such as BOSS (Dawson {\it et al.\/}, 2013), and in the near future there will
be yet more data available, from surveys such as big-BOSS and also potentially from ASKAP and Aperitif
in the radio (Duffy, {\it et al.\/}, 2012), to strengthen this test of fundamental physics.

\deltext{ZPL suggested that, in principle, one could use the difference in redshift offsets observed
for X-ray gas, assuming that can be done, and measurements of galaxies as a probe of the anisotropy
of the velocity dispersion tensor in clusters.  The analysis here shows that the strength of the
kinematically sourced redshifts depends on the luminosity weighting scheme adopted, whereas, to the
extent that the shape of the luminosity function is independent of position, this would not bias
the measurement of the gravitational redshift.  This provides, in principle, another
way to constrain the orbital anisotropy.}

\section{Acknowledgements}

The author enjoyed stimulating discussions on this subject
with Bob McLaren, \newtext{Pat Henry,} Harald Ebeling, William Burgett, Bill Unruh, Richard Ellis, 
and Alex Szalay, and is particularly grateful
for useful input from Scott Tremaine and John Peacock.  
This study was aided by the support of Cifar.

\appendix
\section{Predicting the redshift distribution}

We have estimated above the effects on the mean redshift.  However,
what is actually measured is not a simple centroid, since there
are foreground and background galaxies so what WHH did was to fit the
distribution of redshifts relative to the BDG to a model with a background consisting of a 
linear ramp and a cluster consisting of a double Gaussian.  Given
a theoretical model for the PSD, either analytic or obtained by
stacking clusters found in a simulation, one would like to generate
the predicted distribution of redshifts as a function of impact
parameter.  A convenient way to do this is to note that the
observed redshift $z$ expressed as a recession velocity is, as before,
$\beta'_x = -z = \beta_x - \beta_x^2/2 - \beta_\perp^2 / 2 + \Phi/c^2$,
where we have now separated $\beta^2$ into line of sight and transverse
components.  Thus $d \beta'_x = (1 - \beta_x) d \beta_x$; i.e. the Jacobian
of the transformation from velocity $\beta$, with respect to the rest-frame
observers, to measured redshift $\beta'$ is $1 - \beta_x$.  Conservation of
particles requires that the observed density of particles as a function
of position, redshift and transverse velocity $\rho'(\br, \beta_x', \bbeta_\perp)$
satisfies $\rho'(\br, \beta_x', \bbeta_\perp) d\beta_x' = \rho_{\rm LC}(\br, \beta_x, \bbeta_\perp) d\beta_x$ so
\begin{equation}
\rho'(\br, \beta_x', \bbeta_\perp) = \rho_{RF}(\br, \beta_x' + \beta_x^2/2 + \beta_\perp^2 / 2 - \Phi/c^2, \bbeta_\perp)
\end{equation}
i.e.\ the density of objects in position, radial and transverse velocities is a mapping of the 
rest-frame PSD with a displacement along the $\beta_x$ axis.  Note that $\beta_\perp^2$ here
denotes the sum of the squares of the two transverse velocity components.

We can now expand the RHS as a Taylor series for small displacement.  We also want to
evaluate this at $t = x/c$, which we can also treat as a small displacement, resulting in
\begin{equation}
\begin{split}
\rho'(\br, \beta_x, \bbeta_\perp) & = \rho(\br, \beta_x, \bbeta_\perp) \\
& + (\beta_x^2 / 2 + \beta_\perp^2 / 2 - \Phi / c^2) 
\frac{\partial \rho(\br, \beta_x, \bbeta_\perp)}{\partial \beta_x} \\
& + \frac{x}{c} \dot \rho(\br, \beta_x, \bbeta_\perp)
\end{split}
\end{equation}
where dot denotes partial derivative with respect to time, and we have dropped the prime on $\beta_x$.
Integrating over the transverse velocity components gives
\begin{equation}
\begin{split}
\rho'(\br, \beta_x) & = \rho(\br, \beta_x) \\
& + (\beta_x^2 / 2 + \langle \beta_\perp^2 \rangle / 2 - \Phi / c^2) 
\frac{\partial \rho(\br, \beta_x)}{\partial \beta_x} \\
& + \frac{x}{c} \dot \rho(\br, \beta_x).
\end{split}
\end{equation}
As a sanity check, if we ignore the last term, multiply by $\beta_x$, and integrate over space and velocity,
assuming $\rho$ to be an even function of its arguments, we find
$\delta z = \langle -\beta_x \rangle = \langle \beta_x^2 \rangle + \langle \beta_x^2 + \beta_\perp^2\rangle / 2 - \Phi / c^2$
in accord with equation (5).

We could integrate this expression over line of sight distance to get the distribution function for the
observed redshift as a function of the impact parameter, but that would not properly allow for the
fact that we observe in a cone, nor would it incorporate the surface brightness boosting effects.  Both
of these can be allowed for simply by multiplying the first term on the RHS by the factors
$1 + (3 + \alpha(Z)) \delta(Z) \beta_x$ and $1 + 2 H x (\delta(Z) - 1) / cZ$.\comment{Sign error corrected here and below.}
Linearising the result
gives:
\begin{equation}
\begin{split}
\rho'(\br_\perp, \beta_x) & = \rho(\br_\perp, \beta_x)
+ \int dx \left\{ \right.\\
& (\beta_x^2 / 2 + \langle \beta_\perp^2 \rangle / 2 - \Phi / c^2) \frac{\partial \rho(\br, \beta_x)}{\partial \beta_x} \\
& + ( (3 + \alpha(Z)) \delta(Z) \beta_x + 2 H x (\delta(Z) - 1) / cZ) \rho(\br, \beta_x) \\
& + \frac{x}{c} \dot \rho(\br, \beta_x)\left.\right\}.
\end{split}
\end{equation}

This is valid for an individual cluster.  If we average over the population of clusters and denote averaged properties
as e.g.\ ${\overline \rho} = \int dC P(C) \rho / \int dC P(C)$ then we have
\begin{equation}
\begin{split}
{\overline \rho}'(\br_\perp, \beta_x) & = {\overline \rho}(\br_\perp, \beta_x)
+ \int dx \left\{ \right.\\
& \frac{\beta_x^2}{2} \frac{\partial {\overline \rho}(\br, \beta_x)}{\partial \beta_x}
+ \frac{\partial}{\partial \beta_x}\overline{(\langle \beta_\perp^2 \rangle / 2 - \Phi / c^2) \rho(\br, \beta_x)}\\
& + ( (3 + \alpha(Z)) \delta(Z) \beta_x + 2 H x (\delta(Z) - 1) / cZ) {\overline \rho}(\br, \beta_x) \\
& + \frac{x}{c} \dot {\overline \rho}(\br, \beta_x)\left.\right\}.
\end{split}
\end{equation}

\newtext{The above formulae are valid in the limit $Z \ll 1$.  The extension to
higher redshift is straightforward: Replace $1/Z \rightarrow d \ln D_A / d \ln (1+Z)$.  However
this is only valid if galaxies are assumed to be non-evolving; in reality evolution will
introduce corrections of similar magnitude.}

With an ensemble average cluster PSDF, along with the average of this
times $\langle \beta_\perp^2 \rangle / 2 - \Phi / c^2$  from e.g.\ a cosmological simulation, this
expression, after integrating along the line-of-sight, 
provides the predicted distribution function for the observed redshifts which can then be analysed
in precisely the same way at the real data (e.g.\ finding the shift of the velocity distribution by modelling) to
obtain the predicted redshift offset as a function of impact parameter.  This would also 
allow comparison of predicted and observed higher order moments of the velocity distribution such
as skewness and kurtosis.

\end{document}